\documentstyle[12pt]{article}
\begin{document}

\overfullrule 0 mm
\language 0
\centerline { \bf{ CLASSICAL ELECRODYNAMICS:}}
\centerline { \bf{ PROBLEMS OF RADIATION REACTION}} 
\vskip 0.5 cm
\centerline {\bf{  
Alexander A.  Vlasov}} 
\vskip 0.3 cm
\centerline {{  High Energy and Quantum 
Theory}} \centerline {{  Department of Physics}} \centerline {{  
Moscow State University}} \centerline {{  Moscow, 119899}} 
\centerline {{  Russia}}
\vskip 0.3 cm
{\it  
There are known problems of Lorentz-Dirac equation for moving with 
acceleration charged particle in classical electrodynamics. The 
model of extended in one dimension particle is proposed and shown 
that electromagnetic self-interaction can lead (with appropriate 
choice of retarded and advanced interactions) to zero change in 
particle momentum.  The hypothesis is formulated:  all relativistic 
internal forces of various nature can give zero change in particle 
momentum}

03.50.De
\vskip 0.3 cm
{\bf {0.}}
\vskip 0.3 cm
Since the famous Dirac's paper on relativistic radiation reaction in
classical electrodynamics, many textbooks and research articles were
published on that theme. Among them are [1-11], where one can find 
the discussion of the related problems: mass renormalization and 
its nonuniqueness, runaway solutions and the use of the advanced 
interaction.( These problems of radiation reaction one can find 
also in other classical field theories - scalar field theory, 
gravitational theory, etc.)

First of all we give brief review of some problems (sections 1-3) 
and then in section 4 consider the model of an extended particle.
\vskip 0.3 cm
{\bf {1.}}
\vskip 0.3 cm
It is well known that the x-component of electric field $\vec E$ 
produced by point particle with charge $e=1$ moving along x-axis is 
( Lienard-Wiechert solution; we take the units $c=1$) $$E= 
{(1+Nv')N \over (1-Nv')(x-R')^2} \eqno(1.1)$$ here $R'=R(t')$ - is 
the  trajectory of particle motion and $v'=v(t')= dR'/dt'$ - is the 
particle velocity, $t'$ - is the retarded time:  $t'=t-|x-R'|$, $N$ 
- is the unit vector:  $N=(x-R')/|x-R'|$

To consider self-interaction one must expand field  $E$ near the 
particle. One can do it in different 
ways.  For example:

1. Let $\epsilon=t-t'$  (i.e. $|x-R'|=\epsilon$), $\epsilon \to 0$.
Then
$$ <E>={1\over 
2}\left[\left(E\right)_{N=1}+\left(E\right)_{N=-1}\right]=$$
$${2v\over \epsilon^2 (1-v^2)}-{2\dot v 
(1+v^2)\over\epsilon (1-v^2)^2}+\left[{\ddot v 
(1+v^2)\over (1-v^2)^2} +{(\dot v)^2 2v(3+v^2)\over 
(1-v^2)^3}\right] \eqno(1.2)$$

2. Let $\epsilon=|x-R(t)|$, $\epsilon \to 0$.
Then
$$ <E>={1\over 
2}\left[\left(E\right)_{N=1}+\left(E\right)_{N=-1}\right]=$$
$$-{\dot v\over \epsilon}+\left[{2\ddot v 
\over3 (1-v^2)^2} +{(\dot v)^2 2v\over 
(1-v^2)^3}\right] \eqno(1.3)$$

3. Let $\epsilon=|x-R'|\sqrt{{(1-Nv') \over (1+v'N)}}$,   
$\epsilon \to 0$.  Then $$ <E>={1\over 
2}\left[\left(E\right)_{N=1}+\left(E\right)_{N=-
1}\right]=0\eqno(1.4)$$
here dot means differentiation with respect to $t$, all quantities 
in $<E>$ are taken at moment $t$.

Infinite at $\epsilon \to 0$ terms in (1.2-4) lead to mass 
 regularization in equation of particle motion.  Finite  at 
$\epsilon \to 0$ terms in (1.2-4) are what is usually called "the 
radiation force".  Finite terms in (1.2-1.4) 
 disagree each other and only  finite term in (1.3) gives the known 
 expression for Lorentz-Dirac radiation force.
 
Thus different ways of expansion of particle self-field may lead to 
 different "radiation forces". From mathematical point of view it is  
 obvious result - the expansion near infinity is not well defined 
 operation.

This is a one of various aspects of the problem of mass 
regularization and its nonuniqueness (see also [2,17]).
 
  \vskip 0.3 cm {\bf {2.}} \vskip 0.3 cm

Among other problems, the one, which, to my opinion, draws not 
enough attention in the literature -  is the problem of instability 
of preaccelerative "physical" solution of relativistic Lorentz-Dirac 
equation.

Let the point particle with mass $m$ and charge $e$ move under 
some external force $F$ along $x$-axis. The relativistic 
Lorentz-Dirac  equation reads: 
$$ {m \dot v \over (1- (v/c)^2)^{3/2} }= F
+{2e^2 \over 3 c^3 }\left[ {\ddot v \over (1- (v/c)^2)^{2}} +
{3v (\dot v)^2 \over c^2(1- (v/c)^2)^{3}}  \right] \eqno (2.1) $$
here dot means differentiation with respect to $t, \ v=\dot x$.

Take the dimensionless variables $\eta$ and $\tau$,  dimensionless 
force $f$ and dimensionless radiation parameter $\gamma$ and 
 introduce scale multiplies $a,b,p$:  
 $$a \sim x_{cl}, \ b \sim x_{cl}, \ x_{cl}={e^2\over mc^2}, \ 
 p={a\over b}$$ $$x=a\eta, \ ct=b\tau, \  f=F{b^2 \over amc^2}, \ 
 \gamma={2\over 3b}x_{cl}, $$
here $x_{cl}$ - the classical particle radius.

In terms of new variables equation (1) becomes 
  $$ { \ddot \eta \over (1- (p\dot\eta)^2)^{3/2} }= f 
+\gamma\left[ { \dot {\ddot \eta }\over (1- (p\dot\eta)^2)^{2}} +
{3\dot \eta (\ddot \eta)^2 p^2 \over (1- (p\dot\eta)^2)^3}  \right] 
\eqno (2.2) $$ here dot means differentiation with respect to   
$\tau$.

With the help of relativistic "velocity" $u$:
  $$u ={ \dot \eta \over (1- 
(p\dot\eta)^2)^{1/2} } ={ v/(cp)\over (1- 
(v/c)^2)^{1/2} } \eqno(2.3)  $$ 
 "acceleration" $w$:  
$$w={du \over d\tau} \eqno(2.4)$$ and dimensionless proper time $s$:  
 $${ b \over c(1- (v/c)^2)^{1/2} } {d \over 
dt}=(1+ (p u)^2)^{1/2} {d \over d\tau}= {d\over ds} $$
equation (2.2) can be put in more simple form:
$$w=f +\gamma {dw \over ds} \eqno (2.5)$$ 
The "solution" of (2.5) is obvious:
  $$w=w(s)=-{1\over \gamma}\exp{(s/\gamma)} 
\int\limits_{s_0}^{s}dxf(x)\exp{(-x/\gamma)} \eqno(2.6)$$
with "initial" value $w_0$:
$$w_0=\left({ \ddot \eta \over (1- (p\dot\eta)^2)^{3/2} 
}\right)_{s=0}=
{1\over \gamma}\int\limits_{0}^{s_0}dxf(x)\exp{(-x/\gamma)} $$

Integration of (2.4), taking into consideration  (2.3) and  (2.6), 
yields the following "solution" (strictly  speaking, the integral 
equation) for particle velocity $v$:  $${1\over p}\ln { \sqrt{ 
 {(1+v/c)(1-v_{0}/c) \over (1-v/c)(1+v_{0}/c)} } }=$$ 
$$\int\limits_{0}^{s}dzf(z) 
-\exp{(s/\gamma)}\int\limits_{0}^{s}dzf(z)\exp{(-z/\gamma)}
+ \gamma w_0\left(\exp{(s/\gamma)}-1\right) \eqno(2.7) $$
here $v_0$- the "initial" velocity.

The form  of Lorentz-Dirac equation  similar to (2.7) was given in 
 [2,3,4,7]. Our form (2.7) is 
 convenient for analysis -  from it immediately follows that:

(i) The peculiar features of Lorentz-Dirac equation do not 
qualitatively depend on scale multiplies $a,\ \ b,\ \ p$,  so these 
features are valid as for "small", so for "large", "classical" 
distances.

(ii) If  conditions of a problem permit to consider the limit $s\to 
\infty$ (all integrals in (2.7) are not divergent for $s<\infty$ ) 
then all "solutions" (2.7) must be "runaway" - $|v|\to c$, with one 
exception.

(iii)  This exception is the particular case of zero asymptotic 
value of "acceleration"  $w$:  $s_0= \infty$ in (2.6) and
$$w_0={1\over \gamma} 
\int\limits_{0}^{\infty}dxf(x)\exp{(-x/\gamma)} \eqno (2.8)$$ 
With (2.8), the R.H.S. of (2.7) takes the form
 $$\int\limits_{0}^{s}dzf(z) 
+\exp{(s/\gamma)}\int\limits_{s}^{\infty}dzf(z)\exp{(-z/\gamma)}
- \int\limits_{0}^{\infty}dxf(x)\exp{(-x/\gamma)} \eqno(2.9) $$
 Then for $s\to \infty$ and for "well defined" force $f$ the 
 velocity $v$ does not reach $c$. But the price for this is
the preacceleration and backward in time 
integration (see equation (2.9) ) with accompanying paradoxes 
(some of them are discussed in [7,8,9]) (see also section 3.).

(iv) In literature the "solution" (2.9) is often called  
"physical", but from equation (2.7) it is easy to see that (2.9) is 
unstable under small deviations of "acceleration" $w$ from zero 
value at infinite "future": due to (2.7) these initially small at 
$s=+\infty$ deviations $\delta$ grow at least as $e^{\delta}$.

The instability of the "physical" solution one can also verify with 
the help of numerical calculation: the instability was found in 
[10] and was checked by author for the force $f$ 
formed by external charge at rest.

Following (iv) one can state that there are no stable "nonrunaway" 
solutions of Lorentz-Dirac equation, at least in one-dimensional 
case.

\vskip 0.3 cm
{\bf {3.}}
\vskip 0.3 cm

Here we present one more paradox of preacceleration: the
formation of unavailable area of initial data, i.e. the formation of
"event" horizon, absent in classical equation without radiation
reaction.

Consider for simplicity the nonrelativistic case (with units $c=1$).

Let the point particle with mass $m$ and charge $e$ move (following
the classical nonrelativistic equation without radiation reaction)
under the influence of the external force $F$ along the $x$-axis:
$$ {d^2 x \over dt^2} ={F(x)\over m} \eqno(3.1)$$
Let's take the force $F(x)$ in the form of a positive  step (this 
form of the force is often used in the literature - for ex., [2,3]):  
$$ F =\left (\matrix{0, &x<-x_{0} \cr mA &-x_{0}<x<0\cr 0, 
&0<x\cr}\right ) \eqno(3.2)$$ here $A>0,\ \ \ x_{0}>0$.  In the 
method of backward integration one must take the "final" data (i.e.  
for $t\to \infty$) and zero final acceleration.  Thus for (3.2) we 
assume free motion of the particle in the future with velocity $v$:  
$$x=vt \eqno(3.3)$$

Suppose that the point $x=0$ is achieved at $t=0$, and the point
   $x=-x_{0}$ - at  $t=t_{0}<0$ (the value of $t_{0}$ we shall find
below).

Then the integration of (3.1-2,3) with appropriate boundary conditions
 (for the second order equation (3.1) the position and the velocity of
  particle must be continuous) yields:
   $$ x =\left (\matrix{vt, &t>0 \cr vt+At^2/2
&t_{0}<t<0\cr ut+b, &-\infty<t<t_{0}\cr}\right ) \eqno(3.4)$$

Values of $u,\ b,\ t_{0}$ are determined from the
matching conditions: $$u= \sqrt{v^2-2x_{0}A},\ \ t_{0}={-v+u \over
A},\ \ b=-x_{0}-ut_{0}\eqno(3.5)$$
Following (3.4) in the regions free of force ($x>0$ and $x<x_{0}$ 
) the particle motion is free.

Consider the   case of small velocity ${dx \over dt}=u$ at
the point $x=-x_{0}$:  $$u=\sqrt{v^2-2x_{0}A}=\epsilon,\ \ \  
\epsilon \to
 0\eqno(3.6)$$
We see that though the value of $u$ is small, the
particle can reach from the future all points on the $x$-axis:
$-\infty<x<+\infty$.

Consider now the same particle with the same final data moving under
 the same external force $F$ (3.2) but obeying the nonrelativistic
Lorentz-Dirac equation
 $${d^2 x \over dt^2}-k{d^3 x \over dt^3}
={F(x)\over m} \eqno(3.7)$$ here $k$ ($k\approx
x_{cl}$-the classical radius of a particle, $k>0$) and the second
term on L.H.S. of (3.7) deal with the radiation force.

As the equation (3.7) is of the third order, the acceleration of
the particle also must be continuous. Then the solution of the above
problem for equation (3.7) with appropriate boundary conditions yields
(the point $x=0$ is achieved at $t=0$, and the point
   $x=-x_{0}$ - at  $t=t_{1}<0$ )

  $$ x =\left
  (\matrix{vt, &t>0 \cr (v+kA)t+At^2/2
+k^2A(1-\exp{(t/k)})&t_{1}<t<0\cr ut+b+c\exp{(t/k)},
&t<t_{1}\cr}\right ) \eqno(3.8)$$

Values of $u,\ b,\ c,\ t_{1}$ are determined from the
matching conditions:
  $$-x_{0}=ut_{1}
+b+c\exp{(t_{1}/k)}\eqno(3.9a)$$
$$-x_{0}=(v+kA)t_{1}+A(t_{1})^2/2
+k^2A(1-\exp{(t_{1}/k)})\eqno(3.9b)$$
$$v+kA+At_{1} -kA\exp{(t_{1}/k)})=
u+(c/k)\exp{(t_{1}/k)}\eqno(3.9c)$$
 $$A-A\exp{(t_{1}/k)}
=(c/k^2)\exp{(t_{1}/k)}\eqno(3.9d)$$
Term with exponent in (3.8) for $x<-x_{0}$ (i.e. "before" the action 
of
the force) describes the effect of preacceleration.

From equation (3.9d) immediately follows that $c>0$ for $t_{1}<0$.

Consider the   case of zero  particle velocity ${dx \over
dt}$ at the point $t=t_{1}<0,\ \ x=-x_{0}$:
 $${dx \over dt}=u+(c/k)\exp{(t_{1}/k)}=0 \eqno(3.10)$$
In (3.10)   the
value of $u$ must be negative because the value of   $c$ is
positive.

Equation (3.10) with the help of the system (3.9) can be rewritten as
  $$z^2/2-1-p= \exp{(z)} (z-1)
\eqno(3.11)$$ here $z=t_{1}/k<0,\ \ p=x_0/(k^2A)>0$.
For our goal it is important to note that the equation (3.11) always
has  solution for $z<0$ and this solution varies weakly with
small changes in parameters of the problem under consideration.

Consequently if we consider the similar to (3.6)  case
of small particle velocity  at the point $x=-x_{0}$:  $${dx \over
dt}(t_2)=\epsilon,\ \ \ \epsilon \to 0\eqno(3.12)$$
then $t_2 \approx t_1$ and, as for (3.10),
  $$u<0,\ \ \ c>0 $$.  These inequalities lead to the conclusion 
that the particle velocity in the free of force region ($x<-x_{0}$) 
- ${dx \over dt}=u+(c/k)\exp{(t/k)}$, with positive value $\epsilon$ 
at $x=-x_{0}$, must inevitably take (with time decrease) zero value 
at some moment $t=t_{min}<t_2$ and some point $x=x_{min}<-x_{0}$.

So all $x$ on the left of $x_{min}$: $x<x_{min}$, become unavailable
on contrary to  solution (3.6) of the  equation (3.1),
where all $x$-axis  is available for the particle. Thus the event
horizon is formed: not all initial data, physically possible from the
point of view (3.1), can be achieved by backward in time 
integration method.

 This paradox (as all the other) makes unrealizable the desire to 
solve problems of Lorentz-Dirac equation using preacceleration and 
backward in time integration.
\vskip 0.3 cm
{\bf {4.}}
\vskip 0.3 cm

If to accept that there are problems with the Lorentz-Dirac 
equation, then one can try to solve them considering radiation 
forces of another type.

In the literature there are many examples of different forms of 
radiation forces. Let's mention some of them.
\vskip 0.5 cm
{\it 4.1}
\vskip 0.5 cm
Following the works of Teitelboim school [5,6], when a charge is 
accelerated by a external field, it has a bound momentum $P^i$ given 
by
$$P^i=P^i_{part}+P^i_{bound}=mu^i-m\cdot k \cdot {du^i\over 
ds}, \ \ k=2e^2/(3m)$$ 
With it the Lorentz-Dirac eq. must be rewritten as
 $${dP^i\over ds}=F^{i} +m \cdot k u^i{du^n\over ds}{du_n\over ds} 
$$ 
\vskip 0.5 cm
{\it 4.2}
\vskip 0.5 cm

In works of Plass [3] and Rohrlich [1]  was shown that taking into 
account the asymptotic conditions (zero values for acceleration at 
infinity) the Lorentz-Dirac eq. can be rewritten in the form of 
integrodifferential (nonlocal) equation:
$$m{du^i\over ds}(s)=\int\limits_{0}^{\infty} d\alpha 
K^i(s+k \cdot \alpha)e^{-\alpha}$$
$$K^i=F^i- m\cdot k u^iu_n{d^2u^n\over ds^2}$$
\vskip 0.5 cm
{\it 4.3}
\vskip 0.5 cm
One can rewrite the Lorentz-Dirac eq. using the iteration 
procedure considering the radiation term as small one (for recent 
paper see, for ex., [20]):

$$m{du^i\over ds}=F^{i};$$
$$m{du^i\over ds}=F^{i} +m \cdot k({dF^i\over 
ds}-u^iu_n{dF^n\over ds}) =F^i_1 $$
$$m{du^i\over ds}=F^{i} +m \cdot k({dF^i_1\over 
ds}-u^iu_n{dF^n_1\over ds}) =F^i_2 $$
$$... etc$$
\vskip 0.5 cm
{\it 4.4}
\vskip 0.5 cm

One can try to consider general form of the eq. of motion for a 
point particle using the distribution theory (Lozada, [14])
$$(\partial_i T^{ij}, \Phi)=0$$
with $T^{ij}$- the total energy-momentum tensor for a point particle 
in an external electromagnetic field.

From this general eq. under some assumptions one can derive in 
particular the Bonnor's theory [21] with total nonconstant mass of a 
particle
$$m=m(s);\ \ m{du^i\over ds}=F^{i}$$
$${dm\over ds}=m \cdot k{du^n\over ds}{du_n\over ds} $$

\vskip 0.5 cm
{\it 4.5}
\vskip 0.5 cm

One can draw the quantum field theory to these problems and to 
consider a quantum particle coupled to a quantum-mechanical heat 
bath (more, one can find the opinion in the literature that only 
quantum field approach can solve the problems of radiation reaction. 
I do no share this point of view as the only one possible). For such 
a system the description in terms of the quantum Langevin equation 
has a broad application (a  far as I know, brownian model of 
synchrotronically radiating particle was first suggested by  
Sokolov and Ternov [18], for recent developments see , for ex., Ford 
et al [18]). Then in the nonrelativistic case the macroscopic eq., 
describing the particle motion, is

 $$m\ddot x+ \int \limits_{-\infty}^{t} 
dt'\mu(t-t')\dot{ x}(t') =f+F_{ext}$$ 
here $\mu$ - the memory function, $f$ - a random force with mean 
zero. 

For microscopic eq., see, for ex., Efremov [18].

All the above approaches use the retarded form of self-interaction. 
What new can give the consideration of a superposition of retarded 
and advanced self-interactions? For this sake let's take the model 
of extended (in some sense) particle.
\vskip 0.5 cm
{\it 4.6}
\vskip 0.5 cm

Consider the hydrodynamic model of an extended particle [15].

Then the particle is described by the mass density $m \cdot f(t,x)$, 
charge density  $\rho(t,x)$ and current density $j(t,x)$, obeying 
 the continuity equations $$m{\partial f \over \partial 
 t}+m{\partial (v\cdot f) \over \partial x}=0;$$ 
 $${\partial \rho \over \partial t}+{\partial j \over 
\partial x}=0 \eqno (4.1)$$ here $v=v(t,x)$ is the hydrodynamic 
velocity of moving extended particle.

Let the particle move under the external force $F(t,x)$ 
along x-axis.  Then the relativistic equation of its motion reads 
(we choose the units $c=1$):  $$m\int dx f(t,x)\left( {\partial  
\over \partial t}+v{\partial \over \partial x} \right) u =\int dx 
\rho(t,x)E(t,x) + \int dx f(t,x)F(t,x) \eqno (4.2)$$ here $v=v(t,x)$, 
$u=u(t,x)= v/\sqrt{1-v^2}$, $E(t,x)$ -is the electric field, 
produced by moving particle (the Lorentz force is absent in 
one-dimension case under consideration and internal forces give 
zero contribution to total force):

$$E= - {\partial \phi \over \partial x} -{\partial A \over 
\partial t}  \eqno(4.3)$$
and electromagnetic potentials $\phi$ and $A$ are
$$\phi(t,x)=\int dx'dt'{\rho(t',x') \over 
|x-x'|}(a\delta_1+b\delta_2),$$
$$A(t,x)=\int dx'dt'{j(t',x') \over 
|x-x'|}(a\delta_1+b\delta_2),\eqno(4.4)$$
with retarded and advanced delta-functions
$$\delta_1= \delta(t'-t+|x-x'|),\ \ \delta_2= \delta(t'-t-|x-x'|) $$
and $a,b$ -  constants.

(The role of retarded and advanced interactions in electrodynamics 
is vividly described in the textbook [16].)

Substitution of (4.4) in 
(4.3) and integration by parts with the help of eq. (4.1) (taking 
zero values for integrals of exact integrands in $x'$, i.e. $\int 
dx' {\partial \over \partial x'} (\rho\cdot\cdot\cdot)=0$), yields 
$$E(t,x)=\int {dx'dt' \over |x-x'|^2}\left(\rho(t',x') {x-x'\over 
|x-x'|}(a\delta_1+b\delta_2)+j(t',x')(a\delta_1-b\delta_2)\right) 
\eqno(4.5)$$
Similar integration by parts for LHS of (4.2) gives the common result 
$$LHS={dP \over dt},\ \ P=P(t)=m\int dx f(t,x) u(t,x) \eqno(4.6)$$
here $P$ - the particle momentum.
Thus the eq. of motion reads
$${dP \over dt} =F_{self} +F_{ext},$$
$$F_{self}=\int dx \rho(t,x) E(t,x),\ \ F_{ext}=\int dx f(t,x) 
F(t,x) \eqno(4.7)$$ This eq. of motion has no second derivative of 
particle velocity; also there is no need in mass renormalization. 
 
If the extended particle is compact, we can use in (4.5,7) the 
standard expansion in powers of $|x-x'|$ (see, for ex.,[2]):
$$\delta (t'-t+\epsilon |x-x'|)=\sum\limits_{n=0}^{\infty} 
{\epsilon ^n |x-x'|^n\over n! }{\partial^n \over (\partial 
t')^n}\delta(t'-t)$$
with $\epsilon=\pm 1$.

Thus in nonrelativistic case we get the known result:
$$F_{self}=-(a+b)\int {dxdx' \over |x-x'|}\rho(t,x')\rho(t,x)
{\partial v(t,x') \over \partial t}+$$
$${2\over3}(a-b)\int dxdx'\rho(t,x')\rho(t,x)
{\partial^2 v(t,x') \over (\partial t)^2} \eqno(4.8)$$
The first term in (4.8) is considered in literature as 
"-(electrodynamic field mass)$\times$(acceleration)", and the 
second - as radiation reaction force.
 
The total 
change in particle momentum is $$\Delta P= P(\infty)-P(-\infty)=\int 
dt {dP \over dt}$$ and the change in particle momentum due to  
self-interaction is $$\Delta P_{self}=\int dt F_{self} $$ 
Thus
$$ \Delta P =\Delta P_{self}+\int dt F_{ext}    \eqno(4.9)$$
Substitution of (4.5) into (4.7,9) gives
 $${ d  P_{self} \over dt}=F_{self}=\int 
dt'dxdx' {\rho(t,x) \over |x-x'|^2}\cdot$$
$$ \left(\rho(t',x') 
{x-x'\over 
|x-x'|}(a\delta_1+b\delta_2)+j(t',x')(a\delta_1-b\delta_2)\right) 
\eqno(4.10)$$
$$\Delta P_{self} =\int dt \left[RHS\ \  of\ \  (4.10)\right] 
\eqno(4.11)$$

The solution of (4.1) we can write as
$$\rho(t,x)= {\partial \Phi(t,x) \over \partial x},\ \ j(t,x)= 
-{\partial \Phi(t,x) \over \partial t} \eqno(4.12)$$
Then integration by parts in (4.10,11) with the help of (4.12) gives 
the following result:  $$\Delta P_{self}=\int 
dtdt'dxdx'\Phi(t,x)\Phi(t',x'){x-x'\over |x-x'|^4}\cdot$$ 
$$\left[-6(a\delta_1+b\delta_2) 
+6|x-x'|\left(a{\partial \delta_1 \over \partial t'}+b{\partial 
\delta_2 \over \partial t}\right)-2|x-x'|^2\left(a{\partial^2 
\delta_1 \over (\partial t')^2}+b{\partial^2 \delta_2 \over 
(\partial t)^2}\right) \right]\eqno(4.13)$$
 In (4.13) the integrand is antisymmetric under 
transformations $$ t \to t',\ \ t'\to t,\ \ x\to x',\ \ x'\to x$$
if $$a=b \ \ (=1/2)$$.
 Then      the whole integral 
(4.13) has identically zero value:  
$$ \Delta P_{self}=0$$
 So for an extended in one dimension particle the 
total change in particle momentum due to its self-interaction is 
zero (if is taken the half-sum of retarded and advanced 
interactions). 

For $a=b$ the radiation term in $F_{self}$ (4.8) is identically zero.

Similar result holds for energy balance: from equations (4.1,2) one
can derive
$$ {d\over dt}W_{kin}=A_{self}+A_{ext}\eqno (4.14)$$
with
$$W_{kin}=\int dx {m\cdot f\over \sqrt{1-v^2}}$$
$$A_{self}=\int dx (j\cdot E),\ \ \ A_{ext}=\int dx (F_{ext}\cdot v)
$$ and $$\Delta W_{kin}=\int dt {d\over dt}W_{kin}=\int
dt A_{self}+\int dt A_{kin} \eqno(4.15)$$
Substitution of  (4.12) into (4.14) gives for $\int dt A_{self}$
identically zero result in the case $$a=b=1/2$$

 Then the natural question arises - where is the 
 source of radiated energy in this case? - The answer is obvious and 
 comes from eq. (4.9, 15) - the source of radiated electromagnetic 
 energy is the external force work. 

If $a\not= b$ then  $ \Delta P_{self}$ is not zero and more, the 
sign of $ \Delta P_{self}$ is not identically negative - for some 
processes it can be negative and for the other - not. The latter 
case one can consider as "antidamping". (For example, if $\rho$ and 
$j$ have the form of moving "extended rigid" body with velocity 
$v(t)=dR(t)/dt$: $\rho(t,x)=A\exp{[-\alpha(x-R(t))^2]},\ \ 
j=v(t)\rho(t,x),$ $A,\alpha$- const, then for $a=1,\ b=0$ (retarded 
self-interaction) $ \Delta P_{self} >0$ if $v>0$ for all $t$.)

Consequently we see that the problems of radiation reaction are 
connected not with the chosen form of radiation force but rather 
with the form of chosen self-interaction - retarded or/and advanced.
\vskip 0.1 cm
{\it {4.7}}
\vskip 0.1 cm

Let's
postulate in section 4.6 the following relations $$f(t,x)= \int dv
f(t,x,v)$$ $$\rho(t,x)= Q\int dv f(t,x,v)$$ $$j(t,x)= Q\int dv v\cdot
f(t,x,v) \eqno(4.16)$$ with $f(t,x,v)$- the distribution function of
extended in "$v$-dimension" particle. Thus we introduce the
stochastic character of interaction of particle with its
self-electromagnetic field ( with "heat bath" in spirit of works 
[18, 19]).  Distribution function $f(t,x,v)$ must obey the 
continuity equation:  $${\partial f(t,x,v)\over \partial 
t}+{\partial (vf(t,x,v))\over \partial x} + {\partial (\dot v 
f(t,x,v))\over \partial v} =0. $$ With it the equations of sec.4.6, 
describing the system "particle + self-field", take the 
self-consistent nonlinear form [19] and the results of section 4.6 
remain valid with the following interpretation:

{\it the stochastic self-interaction of particle  can give zero
contribution to  total changes in particle energy and momentum}.

\vskip 0.3 cm
{\bf {5}}
\vskip 0.3 cm

In the literature one can  often find the statement that the insert
of $E_{self}$ (or $F_{self}$, $W_{em, self}=\int dV
(E^2+H^2)_{self}/8\pi$, etc.) into equation of particle motion
(4.2,7) is an obvious procedure justifying by the principle of
extremum of action.  Meanwhile the latter is formulated strictly only
for closed systems or for systems, interacting with known external
sources possessing known equations of motion. Systems with
dissipation (with "damping" as a consequence of emission to infinity
some amount of energy) may not obey the principle of extremum of
action.  Consequently the choice of the forms of $E_{self}$ (or
$F_{self}$, $W_{em, self}$, etc.), dealing with dissipation, can be
considered as some supplementory hypothesis.

Consider the equations (4.14) and (4.9) together with the balance
equation of classical electrodynamics:
$$ {d\over dt}W_{em}=- I - A_{self}\eqno (4.14)$$
where $W_{em}=\int dV (E^2+H^2)/8\pi$ - the total energy (with
the self-energy) in the volume $V$ with surface $S$ containing the
moving source of electromagnetic field, and $I$ - is the flux of
Poynting-Umov vector through the surface $S$.

Then the latter can be put in the form
$$ {d\over dt}W_{tot}=- I + A_{ext}$$
with $W_{tot}=W_{em}+W_{kin}$
or
$$ \Delta W_{tot}=- \int dt I + \int dt A_{ext} \eqno (5.1)$$
This equation is verified in experiments. More precisely,  the
R.H.S.'s of this equations - because the explicit value for
$W_{tot}$ is unknown in the case of point charged particles. Nor is
known in the experiments the explicit equation of motion of radiating
charged particle. So one can take for self-interaction the
superposition of retarded and advanced interactions without violating
the main equation (5.1).

Thus we can reformulate the hypothesis of section 4 in the following
form:

{\it For an extended (in some sense) particle the contribution of
particle self-interaction to the total energy-momentum balance of
system "particle+field" depends not on the choice of the form of
radiation force, but rather on the choice of superposition of
retarded and advanced fields, describing self-interaction of
an particle.}

Then in classical electrodynamics (in the classical theories of 
scalar, gravitational, etc. fields) the  use in 
eq.  of particle motion radiation reaction force in Lorentz-Dirac or 
another form is not the inevitable procedure, dealing with the 
fact of radiation,  the radiation of field energy can be 
possible due to the work of some external force with zero value 
for "radiation force".

In favor of this hypothesis speaks the following example from the 
classical theory of a scalar field. Consider the model of a scalar 
field with self-interaction in the form:
$$L={1 \over 8\pi}\partial_{p}\phi\partial^{p}\phi 
+g\cdot n\cdot \phi-m\cdot n,\ \ \partial_{p}(nu^p)=0$$
Consider the self-expansion of a scalar-charged sphere. Then the eq. 
of its motion is determined by the discontinuities of the total 
energy-momentum tensor across the sphere. This tensor is constructed 
in rather simple way due to the spherical symmetry of the problem. 
Then the eq. of motion is self-consistent - there is no need to add 
to it radiation force - radiation effect is automatically taken into 
account if one knows the total tensor on each side of the sphere (it 
is necessary to note that, contrary to classical electrodynamics,  
the expanding scalar-charged sphere in the classical scalar 
field theory do radiate). The numerical analysis of such expansion 
shows [22] that there are the solutions that one can interpret as 
"runaway": $v \to \pm 1$. 

Thus the existence of such solutions does not bound up with the 
forms of radiation force.

 \vskip 2 cm \centerline {\bf{REFERENCES}}

  \begin{enumerate}
  \item
  F.    Rohrlich, {\it Classical Charged Particles}, Addison-Wesley,
  Reading, Mass., 1965.
\item
D.Ivanenko, A.Sokolov,  {\it Classical field theory} (in
russian), GITTL, Moscow, 1949.
A.Sokolov, I.Ternov, {\it Syncrotron Radiation}, Pergamon Press,
    NY, 1968. A.Sokolov, I.Ternov, {\it Radiation from Relativistic
Electron}, AIP, NY, 1986.
\item  
G.Plass, Rev.Mod.Phys., 33, 37(1961).
\item S.Parrott, {\it
Relativistic Electrodynamics and Differential Geometry},
 Springer-Verlag, NY, 1987.

 \item C.Teitelboim, Phys.Rev., D1, 1572 (1970); D2, 
1763 (1970).  \item  E.Glass, J.Huschilt and G.Szamosi, Am.J.Phys., 
52, 445 (1984).  \item S.Parrott, Found.Phys., 23, 
1093 (1993).  \item W.Troost et al.,  preprint hep-th/9602066.  
\item Alexander A.Vlasov, preprints hep-th/9702177; hep-th/9703001.  
\item J.Kasher, Phys.Rev., D14, 939 (1976).  \item S. de Groot, 
L.Suttorp {\it Foundations of Electrodynamics}, North-Holland, 
Amsterdam, 1972 

\item Anatolii A.Vlasov, {\it Statistical Distribution Functions}, 
(in russian), Nauka, Moscow, 1966.
\item Anatolii A.Vlasov, {\it Non-local Statistical Mechanics },  
(in russian), Nauka, Moscow, 1978. 
\item A.Lozada, J.Math.Phys., 30, 1713 (1989).
\item Alexander A.Vlasov,  preprint hep-th/9704072.
\item Anatolii A.Vlasov, {\it Electrodinamica Macroscopica },  
Editora Universitaria, la Habana, Cuba, 1966. 
\item R.Tabensky, Phys.Rev., D13, 267 (1976).
\item A.Sokolov, I.Ternov, Zh. Eksp. Teor. Fiz. v.25, 698 
(1953). G.Ford, J.Lewis and R.O'Connell, Phys.Rev., A44, 4419(1988).
A.Nikishov,  Zh. Eksp. Teor. Fiz. v.110, 2(8),(1996). G.Efremov,  
Zh. Eksp. Teor. Fiz. v.110, 5(11),1629 (1996). \item Anatolii 
A.Vlasov, Zh.  Eksp.  Teor. Fiz. v.8, 291 (1938); Uchenie Zapiski, 
75,n.2, fizika, Moscow State University, Moscow, 1945.
\item 
J.Aguirregabaria, J.Phys.A: Math. Gen., 30, 2391 (1997).
\item
W.Bonnor, Proc.Roy.Soc. London, A337, 591 (1974).
\item
Alexander A.Vlasov, Teor. Mat. Fiz., 109, n.3, 464 (1996).
\end{enumerate}

 \end{document}